# Effect of memory electrical switching in metal/vanadium oxide/silicon structures with VO$_2$ films obtained by the sol-gel method

A. Velichko, A. Pergament, V. Putrolaynen, O. Berezina, G. Stefanovich

Petrozavodsk State University, 185910, Petrozavodsk, Russia

**Abstract.** Electrical switching and rectifying properties of the metal-VO$_2$-Si structures, on both p-type and n-type silicon, with vanadium dioxide films obtained by an acetylacetonate sol-gel method, are studied. The switching effect is shown to be due to the semiconductor-to-metal phase transition (SMPT) in vanadium dioxide. The shift of the switching threshold voltage, accompanied by the memory effect, in forward bias of the p-Si-VO$_2$ anisotype heterojunction is observed. To explain this effect, a model is proposed which suggests the existence of an additional series resistance associated with a channel at the VO$_2$/Si interface, where a SiO$_x$ layer forms during the VO$_2$ deposition process. This resistance is responsible for both threshold switching characteristics, and the memory effect, and the oxygen ion electromigration process is shown to underlie this effect. Potential applications of the observed phenomena, combining the effects of ReRAM and SMPT, in oxide electronics are discussed.

### 1. Introduction

Transition metal oxides (TMO) are of considerable interest for basic research in condensed matter physics and chemistry and, on the other hand, these materials are promising from the viewpoint of the use in various fields of engineering due to their diverse, and often unique, properties [1, 2]. A set of valence states, associated with the existence of unfilled *d*-shells in the atoms of transition metals, leads to formation of several oxide phases with different properties, ranging from metallic to insulating. A number of TMO-based structures possess the property of unipolar or bipolar resistive memory switching which enables implementation of ReRAM (resistance random access memory) devices [3]. On the other hand, many TMOs exhibit semiconductor-to-metal phase transition (SMPT) [2].

Vanadium dioxide is currently considered as one of candidate materials for oxide electronics [1]. It exhibits an SMPT at $T_t = 340$ K and can therefore be used as a material for Mott-transition FETs [1, 2] and memory elements, for thermochromic coatings, optical filters and controllers, sensors, bolometers, sensitive elements for RF and THz radiation control, micro-electromechanical systems, plasmonic modulators, access diodes in oxide ReRAM, etc [2], as well as thermistors [4] and many other optical and electronic devices (see, e.g., [2] and references therein).

Electrical switching associated with the SMPT (or, as this phenomenon is frequently termed, "electrically-triggered SMPT"), has been observed in various MOM and MOS structures with $VO_2$ [5-15]. This switching effect is extremely promising for potential applications in diverse functional devices of oxide electronics, and, thereupon, the study of electrical properties of Si-$VO_2$ heterojunctions is especial importance. However, as is noted in [14], it is a non-trivial task to fabricate $VO_2$ based junctions with rectifying transport properties because of its structural peculiarities, relatively high work function and high carrier density even in semiconducting phase. It should be stressed here that the properties of vanadium dioxide and the quality of the deposited films are strongly dependent on preparation conditions, and hence such a problem as to grow a proper vanadium oxide film on silicon substrate, i.e. a problem of an appropriate deposition technique, is brought to the forefront. There are several works [6-13] where $VO_2$ has been successfully deposited onto silicon substrates, though the authors have either used some buffering sublayers ($SiO_2$, TiN, MgO, Pt) [9-12] or not reported on rectifying *I-V* characteristics of the junctions on pure Si [6-8]. This fact might be accounted for by a relatively intense mutual diffusion of Si and V atoms at the interface [14] which does not allow formation of an abrupt $VO_2$-Si heterojunction. The results on both switching and rectifying properties have been obtained only for the $VO_2$-GaN [15] and $VO_2$-Ge [16] heterojunctions.

In this work we report on electrical switching and rectifying properties of metal-$VO_2$-Si structures, on both p-type and n-type silicon, with vanadium dioxide films obtained by a liquid phase deposition technique (acetylacetonate sol-gel method). A model describing the *I-V* characteristics of the structures is proposed suggesting an additional series resistance associated with a boundary layer at the $VO_2$/Si interface. Possible applications of the observed phenomena in oxide electronics are discussed.

## 2. Experimental details

Thin films of vanadium dioxide were prepared by an acetylacetonate sol–gel method, followed by a finishing annealing in wet nitrogen at 550°C for 35 min with subsequent heating up to 600°C during 5 minutes, onto (100) single-crystalline Si substrates of p-type (Si:B, $\rho = 12$ Ω·cm) and n-type (Si:P, $\rho = 50$ Ω·cm). According to the ellipsometry measurements, the films on p-Si had the thickness of about 175 nm, and those on n-Si – 190 nm. The conductivity jump at the MIT was measured to be about 2 to 3 orders of magnitude.

We emphasize that the sol–gel technique is currently regarded as one of the most promising processes for obtaining thin oxide films, vanadium oxides included [17-20]. In particular, this technique makes it possible to deposit thin-film coatings onto large-area and complex-shape substrates (e.g. by simple spin- or deep-coating [18, 20]) at low temperatures, which, in turn, is of crucial importance for applications in flexible electronics and 3D integrated circuits. Among different alternative routes for the vanadium oxide sol preparation (such as, e.g., the hydrolysis of alkoxides and the technique based on melt quenching), the acetylacetonate chemical method [17, 19] possesses such important advantages as direct formation of the vanadium dioxide phase (without intermediate formation of $V_2O_5$ and subsequent reduction) and convenience of doping by various elements in a wide impurity concentration range, and the latter is often required to control the value of $T_t$. The film preparation procedures, as well as the results on sample characterization, have been presented in more detail elsewhere [17].

The current-voltage characteristics of the junctions were measured with a *Keithley* 2410 SourceMeter using a software package «VA researcher 1.1» [21]. A spring-loaded top contact (gold wire of 40 μm in diameter) was positioned at the surface of vanadium oxide to complete the MOS structure, and the grounded Si substrate served as a bottom contact.

### 3. Results and discussion
### 3.1. Electroforming and switching in structures metal ─$VO_2$─ p-Si

An initial (i.e. when the voltage across the structure is swept first time) *I-V* characteristic is shown in Fig.1. At a negative voltage (curve $I_f$) on the upper (Au metal) electrode, the process of electrical breakdown occurs (region 2-3) and the structure transforms into an intermediate state (curve $II_f$). This process of electrical forming is accompanied by the formation of a switching channel. Then, in case of an increase in the applied voltage, a second breakdown may occur (region 4-5). Thus, there are three stable states in the *I-V* curve ($I_f$, $II_f$ and $III_f$ in Fig.1) due to the described phenomenon of double breakdown, whose physics will be elucidated below. It should be noted that an intermediate state $II_f$ can be switched into state $III_f$ at a slight structure overvoltage.

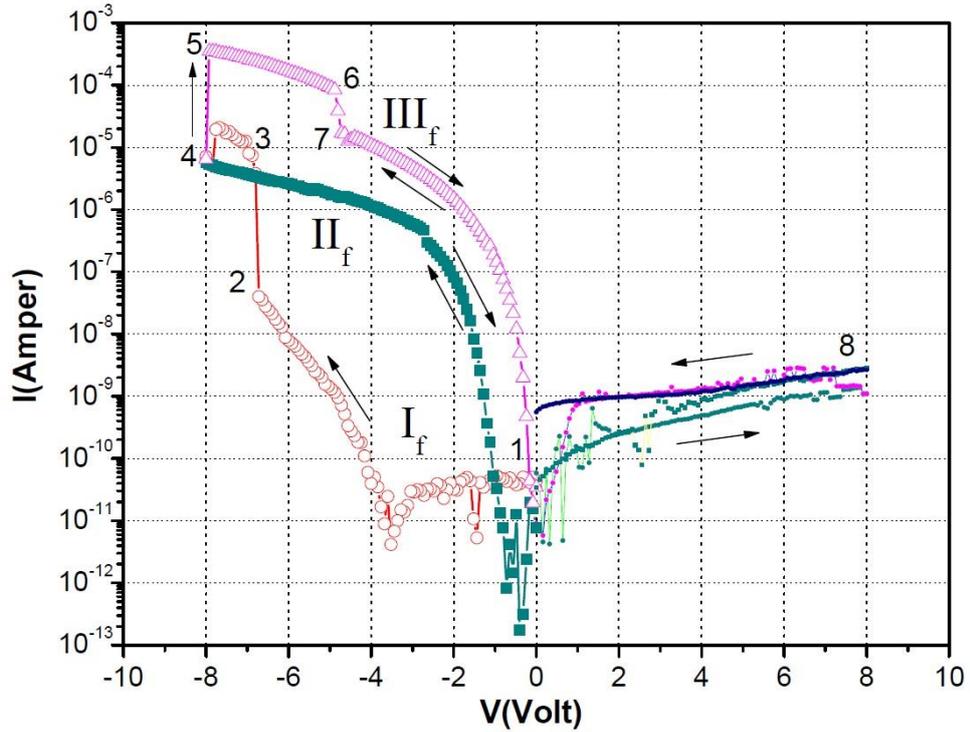

**Fig. 1.** *I-V* characteristic of VO$_2$ on p-Si. I$_f$ – Forming of conductive channel; II$_f$ – Intermediate state of oxide film; III$_f$ – Final state of oxide film; 2-3 Process of electrical breakdown; 4-5 Formation of new conductive channel; 6-7 Semiconductor-metal transition; 1-8 Positive *I-V* curve branch.

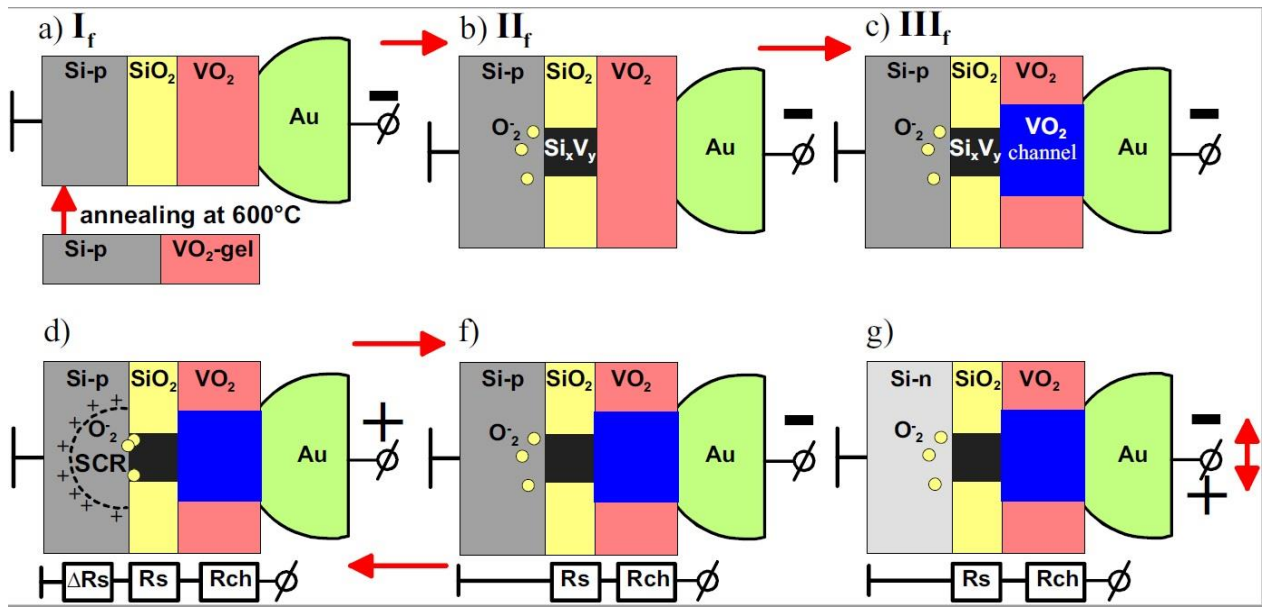

**Fig. 2.** Stages of electroforming of conductive channel in Au-VO$_2$–Si structure. a) Structure after annealing, before electrical forming; b) intermediate state after first breakdown; c) formed structure after second breakdown; d) equivalent circuit at positive polarity; f) that at negative polarity. g) Equivalent circuit of the structure with n-Si (bipolar mode).

The process of multistage forming can be explained on the basis of a three-layer model of the structure depicted in Fig.2. We suppose that the first stage of forming represents in fact the electrical breakdown of a thin silicon oxide layer evidently appearing at the VO$_2$-Si interface during thermal treatment (finishing annealing at 600°C) – Fig.2, a. Thus, the high-resistance state $I_f$ is exactly due to this dielectric layer. Interdiffusion of vanadium, silicon and oxygen at annealing, accompanied by the appearance of boundary layers of $Si_xV_yO$ silicates and Si-O bonds, has been shown in the work [14]. Therefore, one can assume a high resistance $SiO_2$ layer to appear, and an estimate of its thickness from the breakdown voltage (~ 7 V) and critical breakdown field strength (~ $10^7$ V/cm) yields the value of about 7 nm.

Thus, when the voltage is further increased, the breakdown of this $SiO_2$ layer occurs, but the $VO_2$ layer is not affected and the structure turns into state $II_f$ (Fig.2, b). Finally, at an even higher increasing current, forming of the stable switching channel in the vanadium dioxide film occurs (Fig.1, curve $III_f$ and Fig.2, c).

Note also that the existence of stable states at electroforming suggests the possibility of using the structures under study as WORM (write-once-read-many-times) memory cell [22, 23]. A major advantage of this type of memory is a very long storage time, as compared with other types of memory elements, as well as low voltages for electrical switching and reading. Forming of the structure can lead to formation of either a low-resistance channel or a high-resistance one, depending on the type of the film used.

Unlike a conventional WORM memory, which has only two stable states, OFF and ON, the structures studied have three stable states at forming (Fig. 1). This fact suggests that the memory cell made on the basis of the Au-$VO_2$-p-Si structure, is single rewritable. That is, an information unit can be recorded into the cell (Fig. 1, region II), and then this unit can be actively used or stored. However, if needed, information can be overwritten on the storage device (region III), and a data reading interface should differ from a standard one in this case, because it has to distinguish between three cell resistance levels. By analogy with the WORM memory which is usually realized on the basis of polymer [22] or oxide (e.g., IGZO [23]), this $VO_2$-based memory may be called as WTRM – write-twice-read-many-times.

The switching effect (Fig. 3) is associated with an electrothermal instability and the SMPT in this $VO_2$ channel. When a voltage is applied, the $VO_2$ channel is heated up to $T = T_t$ at a certain threshold voltage $V = V_{th}$ (e.g, $V_{th} \approx 6$ V for the negative branch of the *I-V* characteristics – Fig. 3, curve I) and the structure undergoes a transition from an OFF insulating state to an ON metallic state. The switching parameters, particularly, the value of $V_{th}$, are quite stable and do not change during multiple (more than 100 times) cycling if the polarity of the bias is not reversed. A single positive bias (curve III), followed by applying of an opposite polarity voltage, leads to a change of the parameters, viz., to an increase of the threshold voltage (curve II). This state II is metastable, it disappears during one cycling (for about 10 sec.) and the structure turns back into state I.

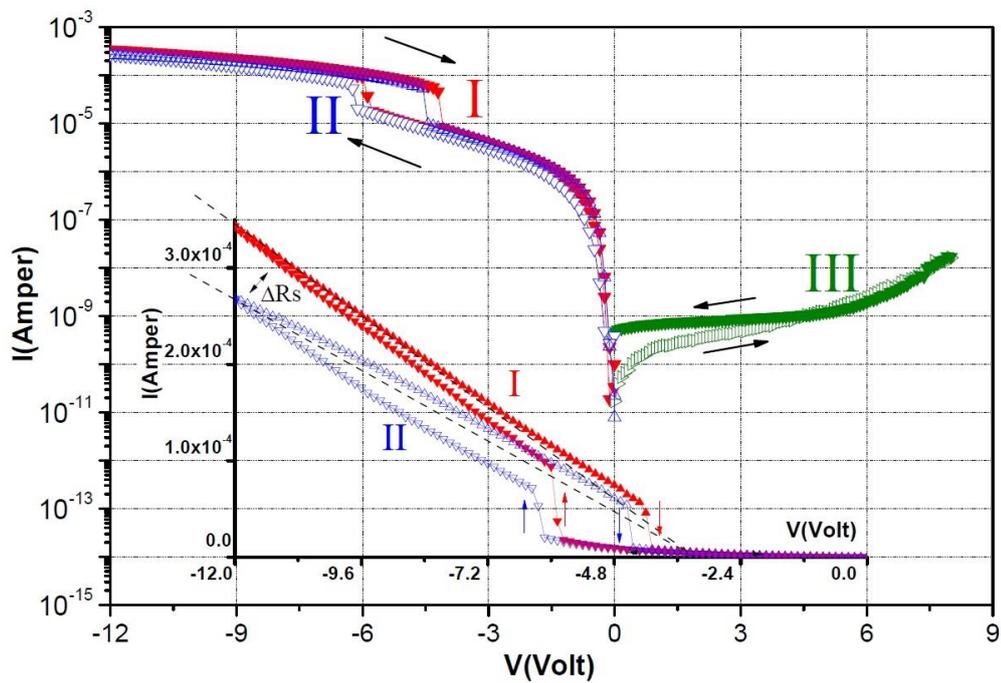

**Fig. 3.** *I-V* characteristic of $VO_2$ on p-Si after final electroforming (inset – the same curve at negative polarity in linear coordinates).

To elucidate a mechanism of the polarity influence on the parameters of the switching effect, as well as the memory effect associated with that, we consider the factors affecting electrical switching due to SMPT. In the model of critical temperature [2, 5], the higher the transition temperature $T_t$, the higher the threshold voltage $V_{th}$ *ceteris paribus*. Also, $V_{th}$ depends on the channel length (film thickness) and radius. Another aspect affecting the value of $V_{th}$ might be the presence of a series resistance $R_s$ in the structure which is not related to the switching channel volume (see Fig.2, f). As the value of this $R_s$ increases, $V_{th}$ increases too, while the threshold current $I_{th}$ should remain constant.

To describe the observed *I-V* characteristics, we assume that the breakdown of a thin $SiO_2$ layer is accompanied by formation of a channel consisting of Si doped with vanadium

($Si_xV_y$), which would presumably possess the n-type conductivity; the assumption of p-type conduction of this channel does not change anything in the subsequent speculations, since the size of this $Si_xV_y$ region is considerably smaller than the space charge region (SCR) width $d_p$ in the p-Si substrate (as estimated in [24], $d_p \gg 50$ nm) where the voltage drops predominantly at positive polarity. This leads to the occurrence of a p-n heterojunction between p-Si and n-$Si_xV_y$ and causes the rectifying behavior of the *I-V* characteristics after electroforming. The n-$Si_xV_y$/$VO_2$ junction has ohmic conductivity, since vanadium dioxide is usually regarded as an n-type semiconductor [15]. After the second forming stage of the as-prepared film and formation of a crystalline $VO_2$ channel, the effect of electrical switching associated with the SMPT [5-15] on the negative branch of the *I-V* curve is observed (Fig.3).

As was mentioned, the switching parameters such as threshold voltages and currents may depend not only on the $VO_2$ channel size and $T_t$, but also on the presence of an additional series resistance $R_s$, $Si_xV_y$ channel in our case. This resistance can vary upon the polarity change due to electromigration of oxygen ions (oxygen vacancies) in the system – the effect akin to the mechanism responsible for the bipolar memory in a number of oxide structures [25].

At positive polarity, the resistance of the channel in $SiO_2$ increases by $\Delta R_s$ (seeFig.2.d), which affects the shift of the threshold characteristics and current reduction in the ON state at the negative polarity (an estimate from the *I-V* curve yields $\Delta R_s \sim 6$ k$\Omega$, see Fig.3, inset).

We now discuss why the $Si_xV_y$ channel resistance is increasing at positive polarity. It is reasonable to assume that, if the channel forming in $SiO_2$ occurs at negative polarity, the elimination of this channel, i.e. its "shutting", should take place at reverse polarity. During the process of forming, a negative voltage corresponds to the forward bias of the p-Si/n-$Si_xV_y$ junction and, at that, oxygen ions transfer into the p-Si (Fig.2, c). Oxygen diffusion into the Si substrate during breakdown of ultrathin $SiO_2$ has also been observed in [26]. At positive polarity, the whole voltage drops across the SCR of the p-Si/n-$Si_xV_y$ junction where there is a high electric field initiating electromigration of oxygen ions toward the $Si_xV_y$ channel which leads to its partial oxidation and an increase in resistance (Fig.2, d). This OFF state is memorized for a long time, at least for more than 10 minutes, which accounts for the memory effect of the structure.

By the time of the re-application of the negative polarity bias, the $Si_xV_y$ channel resistance is restored to its initial value (Fig.2, f). The recovery of the channel resistance to its initial value occurs for the time of one cycling (curve II in Fig.3), and this process is conditioned by electromigration of oxygen from the $Si_xV_y$ channel into the p-Si substrate. Note that the ON and OFF currents change only slightly which indicates the $VO_2$ channel radius invariability, since the critical current density should preserve. Thus, the observed phenomena have an ion drift nature which corresponds to the models of bipolar ReRAM suggested for other materials [3].

### 3.2. Switching in structures metal ─VO$_2$─ n-Si

The n-Si – VO$_2$ contact represents an isotype heterojunction and, as one can see in Fig.4, the *I-V* curve is slightly rectifying (the SCR is low-resistance), almost ohmic: the forward and reverse currents at *V* ~ 1 V differ by less than an order of magnitude. This behavior could be associated with the fact that the work functions of n-Si and vanadium dioxide (in the semiconducting state) are almost equal and the energy barrier is therefore rather low. Note that the work function of VO$_2$ in the semiconducting state is not a strictly defined quantity because the Fermi level location within the gap depends on many factors such as non-stoichiometry, defects, impurities, etc. We suppose that electrical forming results in formation of an Si$_x$V$_y$ channel similarly to that described in previous subsection (see Fig.2, g).

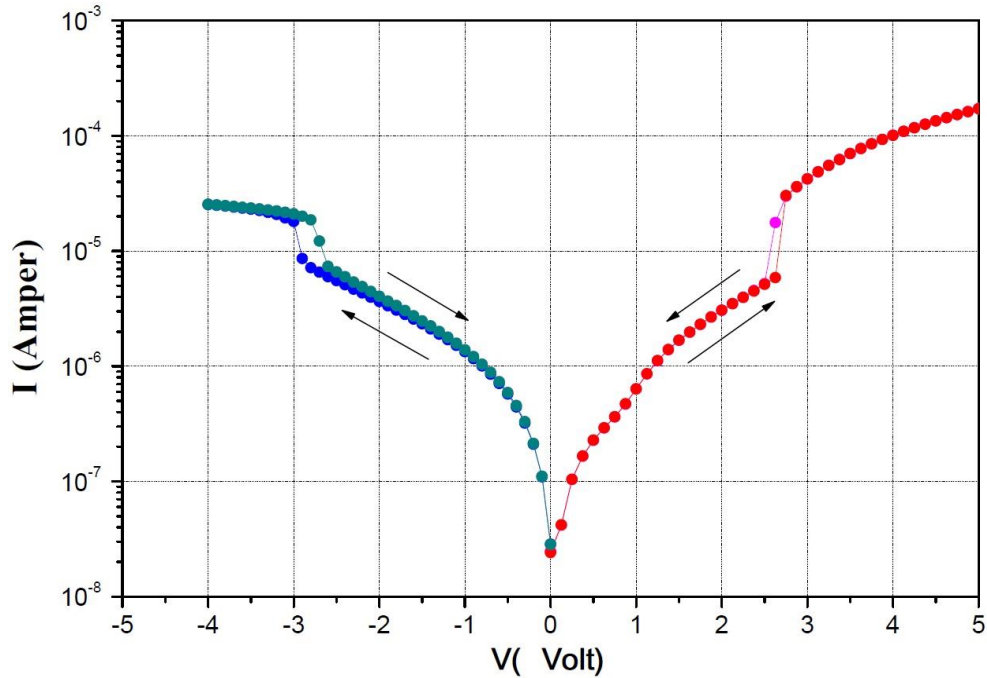

**Fig. 4.** *I-V* characteristic of VO$_2$ on n-Si after final electroforming.

After electrical forming similar to that in previous case of p-Si, the SMPT-induced switching effect is observed in both forward and reverse bias. A slight difference in threshold parameters at the positive and negative polarity is due to a voltage drop across the low-resistance junction SCR, which is different in different biases and, inasmuch as the VO$_2$ switching channel is connected in series with the junction, the genuine voltage drop across the channel at switching is the same in both biases.

A more striking difference is observed for the ON states (at *V* > *V*$_{th}$) of the *I-V* curve at positive and negative polarities. The point is that the work function of metal VO$_2$, ~ 5.3 eV [15,

27], exceeds significantly the electron affinity of silicon, which ensures the Shottky barrier (see Fig.5) at the interface and rectification at $V > V_{th}$. This is exactly what is observed.

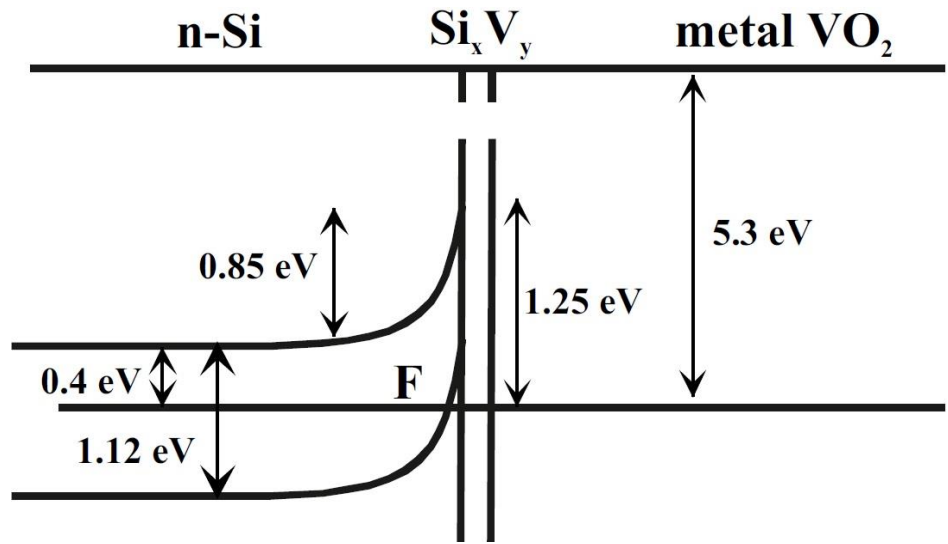

**Fig. 5.** Energy band diagram of "n-Si/metallic VO$_2$" structure. (the Si$_x$V$_y$ region is considerably thinner than the VO$_2$ film thickness and hence it does not affect the band diagram).

When using the n-type Si substrate, the SCR is of low-resistance, no high electric field appears which, in turn, does not allow the electromigration of oxygen ions, the memory effect, and the variation of the switching threshold parameters.

Next, we would like to discuss the results obtained in [13] which are of interest to compare with the results of the present study. A persistent conductivity increase has been demonstrated in vanadium dioxide thin films grown on single crystal silicon by irradiation with swift heavy ions [13]. As one of the explanations, the authors surmise that this conductivity change could be due to the ion-induced modification of a high resistivity interface layer formed during film growth between the vanadium dioxide film and the n-Si substrate, i.e. that the swift heavy ions may generate conducting filaments through this layer [13]. In the present work, such conducting channels are formed at electroforming. The absence of rectification in the I-V curves presented in the work [13] can be explained by the use of n-type Si which is in agreement with our results (Fig.4). Note also that the equivalent circuit presented in [13] supposes the presence of a resistor associated with the above-mentioned interface SiOx layer which is consistent with our assumption of an additional series resistance associated with a channel at the VO$_2$/Si interface. Finally, we would like to emphasize that swift heavy ion irradiation might be an alternative, with respect to electrical forming, route for fabrication of diverse devices for oxide electronics.

## 4. Conclusion

Vanadium dioxide thin films are obtained by the acetylacetonate sol-gel method on silicon substrates. Three distinct effects are observed, viz., electroforming-induced WTRM memory, MIT-induced switching, and bipolar memory switching. Electrical switching is observed in metal-$VO_2$-p-Si and metal-$VO_2$-n-Si structures. The switching effect is associated with the semiconductor-to-metal phase transition in the vanadium dioxide channel which is formed in the as-prepared structure during the process of electroforming. In addition, the structures on p-type silicon demonstrate the bipolar memory switching effect due to the ion transfer.

It is noteworthy that the *I-V* curves of the structures display the diode properties, and the forward and reverse currents differ by three to five orders of magnitude (Fig. 1, points 5 and 8) which means that the memory cell power consumption is negligible at positive polarity and when switching into another resistive state.

In this paper we do not present a detailed study of the memory effects, write and erase modes, optimization of the structure under study, and such a study still remains to be done. The basic idea of the present study is to combine the effects of bipolar memory and electrical switching, which allows the switching threshold parameters alteration at a minor change in the system internal resistance. This opens the opportunity to significantly increase the efficiency of cell status reading, because the ON and OFF currents at switching may differ by several orders of magnitude

Summarizing, the discussed prototype devices, namely, $VO_2$ switches on Si, Si-$VO_2$ bipolar memory, as well as the above described WTRM memory cells, appear to be fairly promising from the viewpoint of oxide electronics perspectives [1, 2].

**Acknowledgements**

This work was supported by the Strategic Development Program of Petrozavodsk State University (2012 – 2016) and the RF Ministry of Education and Science as a base part of state program № 2014/154 in the scientific field, project no. 1704.